\documentclass[prl,showpacs,twocolumn,floats,floatfix]{revtex4-1}
\usepackage{graphicx}
\usepackage{color}
\usepackage{hyperref}
\usepackage{dcolumn}
\usepackage{multirow}

\newcommand{\be}{\begin{equation}}
\newcommand{\ee}{\end{equation}}
\newcommand{\bea}{\begin{eqnarray}}
\newcommand{\eea}{\end{eqnarray}}
\newcommand{\bit}{\begin{itemize}}
\newcommand{\eit}{\end{itemize}}
\newcommand{\bfi}{\begin{figure}}
\newcommand{\efi}{\end{figure}}
\newcommand{\bfic}{\begin{figure*}}
\newcommand{\efic}{\end{figure*}}
\newcommand{\bce}{\begin{center}}
\newcommand{\ece}{\end{center}}
\newcommand{\btb}{\begin{tabular}}
\newcommand{\etb}{\end{tabular}}
\newcommand{\bwt}{\begin{widetext}}
\newcommand{\ewt}{\end{widetext}}

\newcommand{\dd}{{\rm d}}

\begin{document}

\title{Effects of nonlinear inhomogeneity on the cosmic expansion with numerical relativity}

\author{Eloisa Bentivegna$^{a,b}$ and Marco Bruni$^{c}$\\~}

\affiliation{
$^a$Dipartimento di Fisica e Astronomia, Universit{\`a} degli Studi di Catania, Via S.~Sofia 64, 95123 Catania, Italy\\
$^b$INFN, Sezione di Catania, Via S.~Sofia 64, 95123 Catania, Italy\\
$^c$Institute of Cosmology \& Gravitation, University of Portsmouth, Portsmouth PO1 3FX, UK}


\begin{abstract}
We construct a three-dimensional, fully relativistic numerical model of a 
universe filled with an inhomogeneous pressureless fluid, starting from initial
data that represent a perturbation of the Einstein-de~Sitter model. We then measure 
the departure of the average expansion rate with respect to this 
homogeneous and isotropic reference model,
comparing local quantities to the predictions of linear perturbation theory.
We find that collapsing perturbations reach the turnaround point much earlier than expected from 
the reference spherical top-hat
collapse model and that the local deviation of the expansion rate from the homogeneous one
can be as high as $28\%$ at an underdensity, for an initial density contrast of $10^{-2}$.
We then study, for the first time, the exact behavior of the backreaction term
${\cal Q}_{\cal D}$.
We find that, for small values of the initial perturbations, this term exhibits a $1/a$ scaling, and that it is negative with a linearly growing absolute value for larger perturbation 
amplitudes, thereby contributing to an overall deceleration of the expansion. 
Its magnitude, on the other hand, remains very small even for relatively large perturbations.
\end{abstract}


\maketitle

Cosmology as a physical theory of the Universe  was  born soon after the formulation of general relativity one hundred years ago~\cite{Ellis:2015lea}, yet the extent to which relativistic nonlinearity may affect  structure formation remains largely unexplored. 
With the  increasing
volume of cosmological data and their precision, more sophisticated modelling is required, and thus it is becoming timely to quantify  these relativistic effects.
The current theoretical framework for cosmology is based on three main ingredients:  a homogeneous and isotropic Friedmann-Lema{\^i}tre-Robertson-Walker
(FLRW) background, relativistic perturbation theory to describe fluctuations in the early universe and at very large scale, and Newtonian methods, notably N-body simulations,  to study the evolution of  fluctuations into the nonlinear regime of structure formation. 
Reconciling this framework with the observations requires the existence of dark components,  cold dark matter (CDM) and a cosmological constant $\Lambda$ or some other form of dark energy. The resulting
standard cosmological model, $\Lambda$CDM, satisfies a vast class of observational constraints, in particular the high precision measurements of the cosmic microwave background anisotropies~\cite{Ade:2015xua}. However, the existence and nature 
of these dark constituents 
are one of the most debated topics not only in modern cosmology, but also in theoretical physics. One  aspect that has been the subject of  intense debate
is the question whether nonlinear relativistic ``backreaction" effects due to formation of  structures may play an important role in the average cosmic expansion~\cite{Buchert:1999er,
Rasanen:2011ki,Buchert:2015iva,Green:2015bma,Kolb:2004am}.

Quantifying the systematic errors involved in the different modelling approximations,
such as the use of Newtonian gravity for structure formation,
is a crucial undertaking if one wishes to interpret correctly the data which will be
produced by the upcoming precision surveys~\cite{Amendola:2012ys,Maartens:2015mra}.
Whilst some approaches have been introduced
to estimate the role of relativistic corrections in $N$-body simulations~\cite{Bruni:2013mua,Thomas:2015kua,Milillo:2015cva,Adamek:2013wja,Adamek:2014xba,Adamek:2015eda},
the only viable avenue to an exact computation of the systematic
errors resulting from the omission of these effects is the direct numerical
integration of Einstein's equation in the corresponding scenarios.
Integrating the equations of general
relativity, possibly coupled to stress-energy sources, is the field of numerical
relativity, a framework strongly motivated by gravitational-wave--source modelling,
but which has, over the years, developed in a number of
parallel areas such as cosmology, mathematical relativity, and modified gravity~\cite{Cardoso:2014uka}.
Some of this work has already been aimed at studying inhomogeneous
cosmologies~\cite{Bentivegna:2012ei,Yoo:2012jz,Yoo:2013yea,Bentivegna:2013jta,Yoo:2014boa}.
While these numerical-relativity studies do not yet aspire to the level of realism
achieved by $N$-body simulations~\cite{Bruni:2013qta,Fidler:2015npa}, they are useful
testbeds to quantify the relativistic effects of nonlinear inhomogeneity on the cosmic expansion.

In this Letter, we integrate Einstein's equation coupled to an inhomogeneous
irrotational pressureless fluid (dust) with a three-dimensional density profile and no continuous symmetries.
We choose initial data corresponding to a perturbed Einstein-de~Sitter (EdS) model, i.e.~a flat FLRW model 
with dust, with the aim of measuring, with no approximations, the departures of the 
fully nonlinear numerical solution from the idealised FLRW background and its perturbations.
On the numerically-generated spacetimes, we measure a number of local and average properties of cosmological
interest, such as the growth of overdensities
and the formation of voids, the inhomogeneous and average expansion rate, and the
backreaction term defined in the averaging framework~\cite{Buchert:1999er}.
The main results of this study are that (i) those perturbations that are large enough 
to collapse stop partaking in the cosmic expansion (i.e.~reach the ``turnaround'' point) much earlier than 
expected from a spherical top-hat collapse model with the same initial density contrast;
(ii) locally, the effects of nonlinear inhomogeneities
can be substantial, leading to a departure from the average expansion rate of over $28\%$ at
the underdensities; 
and (iii) the average expansion rate is hardly affected by the inhomogeneities, 
with a backreaction term which is never larger than $10^{-8}$.

{\it Method.}
We integrate
Einstein's equation and the fluid conservation equation
using a variant of the Baumgarte-Shapiro-Shibata-Nakamura formulation
~\cite{Nakamura:1987zz, Shibata:1995we, Baumgarte:1998te},
along with the Wilson formulation for the hydrodynamical system
\cite{lrr-2008-7} and
the conformal transverse traceless formulation 
for the Einstein constraints~\cite{lichnerowicz:1944,York:1971hw},
an approach already used in cosmological settings \cite{PhysRevD.60.064011,Giblin:2015vwq,Mertens:2015ttp}.
We choose to represent the spacetime in the synchronous-comoving gauge~\cite{1975Landau}, 
popular in cosmological perturbation theory~\cite{Bruni:2013qta},
which corresponds to the Lagrangian coordinates of the observers at rest with the matter.

To integrate this system, we use the Einstein Toolkit~\cite{Loffler:2011ay,*et}, 
a free, open-source community infrastructure for numerical relativity.
In particular, we use the \texttt{McLachlan} 
code~\cite{mclachlan,*kranc} for the evolution of the gravitational variables, 
the \texttt{Carpet}~\cite{carpet} package for handling adaptive mesh refinement, and
the multigrid elliptic solver \texttt{CT\_MultiLevel}~\cite{Bentivegna:2013xna} 
to generate initial data; this is then coupled to a new module which evolves the hydrodynamical equations.
All equations are discretized using fourth-order finite differencing.

The Einstein Toolkit is routinely used for simulations
in relativistic astrophysics, and passes a variety of tests~\cite{Loffler:2011ay}. 
Likewise, as will be presented elsewhere, the new module correctly reproduces 
several exact cosmological models with varying degrees of inhomogeneity.
All results presented are convergent at the correct rate as the grid
spacing is decreased, and we use this fact to extrapolate the continuum solution
of the evolution system, and estimate the error bars resulting from its numerical
integration at finite resolution. These are the quantities that appear in all
plots. 

{\it Perturbations and averaging.}
We recall two approaches commonly used to solve the evolution system
approximately, so that we can compare our solution to these schemes and check that 
we obtain the correct behavior in the appropriate regime. 

For irrotational dust in the synchronous-comoving gauge, the line element can be written 
(with no loss of generality~\cite{1975Landau}) as $\dd s^2 = -\dd t^2 +\gamma_{ij}\dd x^i \dd x^j$, 
where $\gamma_{ij}$ is the spatial metric. 
For spacetimes that are close enough to a FLRW model, one can use perturbation
theory to follow the departures from the exact background solution. In the matter era,
this is the spatially-flat EdS model, with metric $\bar \gamma_{ij}= a(t)^2 \delta_{ij}$, 
where the scale factor $a(t)$ is a solution of Friedmann's equations
\be
\label{eq:fried1}
\frac{\dot {a}^2}{a^2} = \frac{8 \pi \bar \rho}{3} \quad \quad \frac{\ddot {a}}{a} = - \frac{4\pi}{3} \bar \rho\,,
\ee
the dot represents a time derivative, 
and we denote the EdS-background quantities with an overbar.
The matter continuity equation gives $\bar\rho\sim 
a^{-3} $ for the background density. Starting from the inhomogeneous density $\rho$,
one can define the density contrast $\delta = (\rho-\bar \rho)/\bar \rho$; 
its growth in the synchronous-comoving gauge is governed, at first order, by:
\be
\label{eq:denscont}
\delta'' +\frac{3}{2 a}\delta' -\frac{3}{2 a^2}\delta=0\,. \label{deltaeq} 
\ee
The system of (\ref{eq:fried1}) and (\ref{eq:denscont}) is then solved by:
\bea
a(t) &=& a_i \left(\frac{t}{t_i}\right)^{2/3}\,,  \\
\delta(t) &=& \delta_+ a(t) + \delta_- a(t)^{-3/2} \,, \label{delta}
\eea
where $\delta_+$ and $\delta_-$ are the so-called growing and decaying modes. 
We will use these expressions below as a consistency check in the small-perturbation
regime.

Another useful framework is that of cosmological averaging~\cite{Buchert:1999er}, where Einstein's equation
is reduced from a set of partial differential equations for the fields to a set of
ordinary differential equations in time 
for some of their averages over a given spatial region $\cal D$. Defining its volume as
\be
a_{\cal D}^3 = \int_{\cal D} \sqrt{\gamma} \; {\rm d}^3 x\,,
\ee 
where $\gamma$ is the determinant of the spatial metric $\gamma_{ij}$,
one finds that the average scale factor $a_{\cal D}$ satisfies a system similar to Friedmann's (\ref{eq:fried1}), and in particular that
\be
\label{eq:avea}
\frac{{\ddot a}_{\cal D}}{a_{\cal D}} = - \frac{4 \pi}{3} \frac{M_{\cal D}}{a^3_{\cal D}} + \frac{{\cal Q}_{\cal D}}{3}\,,
\ee
where:
\bea
M_{\cal D} &=& \int_{\cal D} \sqrt{\gamma} \, \rho \, {\rm d}^3 x \\
\label{eq:br}
{\cal Q}_{\cal D} &=& \frac{2}{3} (\langle K^2 \rangle_{\cal D}-\langle K \rangle_{\cal D}^2)-2 \langle A^2 \rangle_{\cal D}.
\eea
Here $K$ is the trace of the extrinsic curvature $K_{ij} \equiv -\dot \gamma_{ij}/2$,
$A^2 = A_{ij} A^{ij}/2$, 
$A_{ij}$ is the traceless part of $K_{ij}$, and $\langle \cdot \rangle_{\cal D}$ denotes the average of a field over ${\cal D}$. 
Note that $-K$ represents the local expansion rate, and in the FLRW background $H=-\bar K/3$ is the Hubble  parameter. 
Whilst this setup
is exact, the computation of ${\cal Q}_{\cal D}$ itself requires tensorial quantities
that do not satisfy ordinary differential equations, i.e.~the system of ordinary differential
equations for the averaged quantities is not closed. 
To circumvent this problem,
one typically closes the system with a well-motivated ansatz 
for ${\cal Q}_{\cal D}$.
One can, for instance, calculate its perturbative behavior: at first order, this term
is identically zero, while at second order it scales as ${\cal Q}_{\cal D} \sim a^{-1}$ \cite{Kolb:2004am,Li:2007ny}, but with a coefficient containing only surface terms of the 
averaging volume, which vanish for periodic domains.
Beyond this order, the analytical approach becomes exceedingly difficult.
A main goal of this Letter is to present an exact measurement of this quantity on an inhomogeneous spacetime.

{\it Results.} Our numerical investigation involves the evolution of a cubic domain of coordinate side $L$, 
with periodic boundary
conditions (since we set $G=c=1$, $L$ will serve as the unit in which all other quantities, including mass and time, are measured). 
We discretize this domain with $160^3$ points (running two lower resolutions with $80^3$ 
and $40^3$ to quantify the error bars).
We choose the initial density profile as that of the EdS model at the time when the Hubble horizon $H_i^{-1}=L/4$,
plus a superimposed perturbation of initial amplitude 
$\delta_i$ (varying between  $10^{-6}$ and $10^{-2}$) and comoving wavelength $L$:
\be
\label{eq:densID}
\rho_{i} = \bar \rho_i (1 + \delta_i \sum_{j=1}^3 \sin \frac{2 \pi x^j}{L})
\ee
The ratio $\rho/\bar \rho$ for $\delta_i=10^{-2}$ is shown in Fig.~\ref{fig:dens}.
As $\delta_i$ decreases, we expect to recover a cubic domain of the EdS model. By increasing
$\delta_i$, we should then be able to observe the onset of nonperturbative effects.
\bce
\bfi
\vspace{0.5mm}
\includegraphics[width=0.235\textwidth]{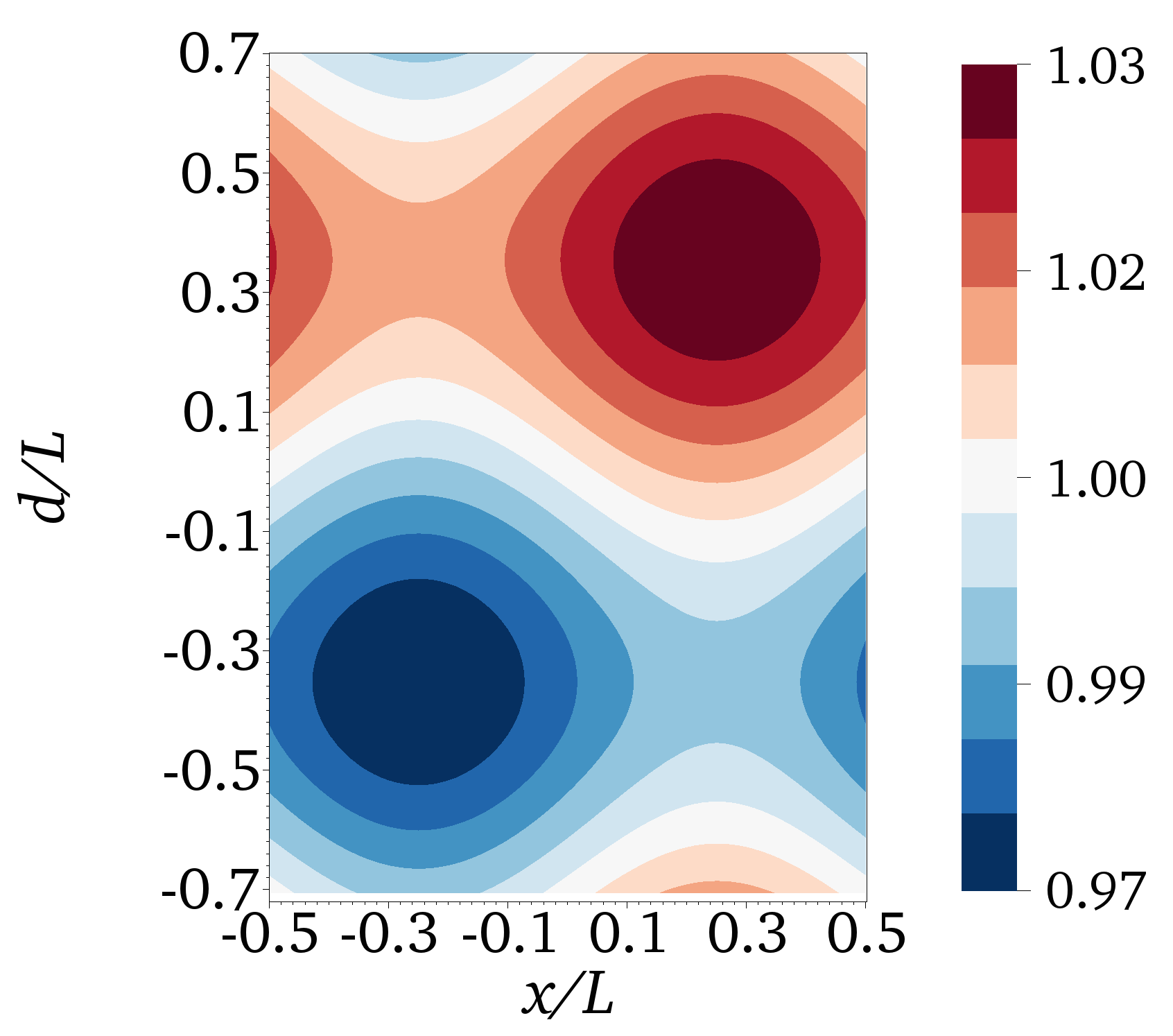}
\includegraphics[width=0.235\textwidth]{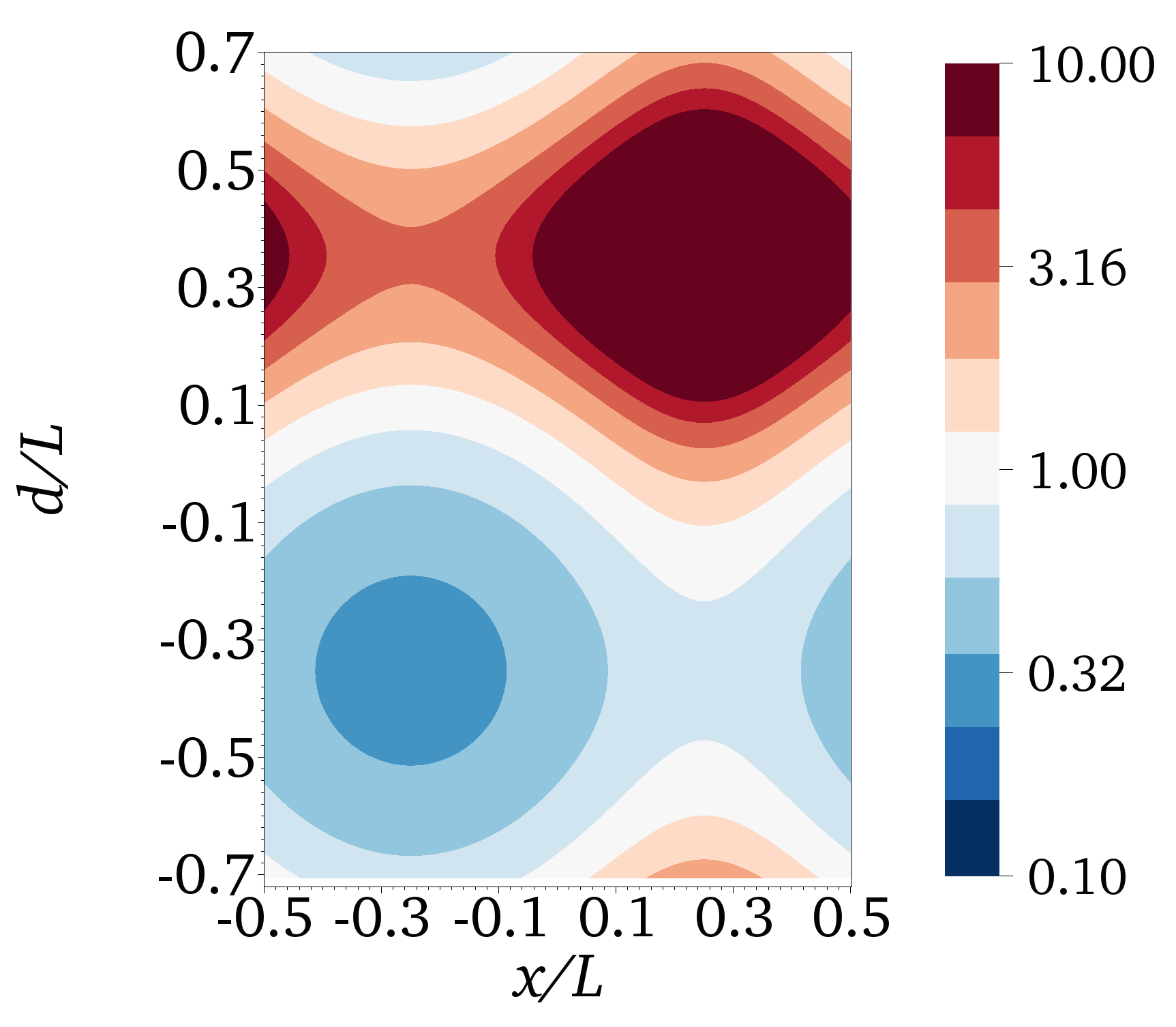}
\caption{
Profile of the matter density ratio $\rho/\bar \rho$ on the $y=z$ plane ($d=\sqrt{y^2+z^2}$) for $\delta_i=10^{-2}$,
when $a = a_i$ (left) and when  $a \sim 96 a_i$  (right).
\label{fig:dens}}
\efi
\ece
We first need to solve the Einstein constraints;
to simplify them, we choose a vanishing traceless part of the 
extrinsic curvature
and a spatially constant $K$.
This corresponds to have, initially, a vanishing first-order perturbation of the expansion and a non-zero decaying mode $\delta_-$  in (\ref{delta})~\cite{Bruni:2013qta}.
The momentum constraint is then identically satisfied, and the Hamiltonian constraint reduces to
the nonlinear elliptic equation:
\be
\Delta \psi - \left (\frac{K_i^2}{12} - 2 \pi \rho_i \right ) \psi^5 = 0
\ee 
where $\psi = \gamma^{1/12}$.
Using (\ref{eq:densID}) and $K_i=\bar K_i=-3H_i=-\sqrt{24 \pi \bar \rho_i}=-12/L$,
we solve this equation with \texttt{CT\_MultiLevel}~\cite{Bentivegna:2013xna}, obtaining the initial profile for  $\gamma$ (normalized to the EdS value) shown in Fig.
\ref{fig:W}.
\bce
\bfi
\vspace{0.5mm}
\includegraphics[width=0.23\textwidth]{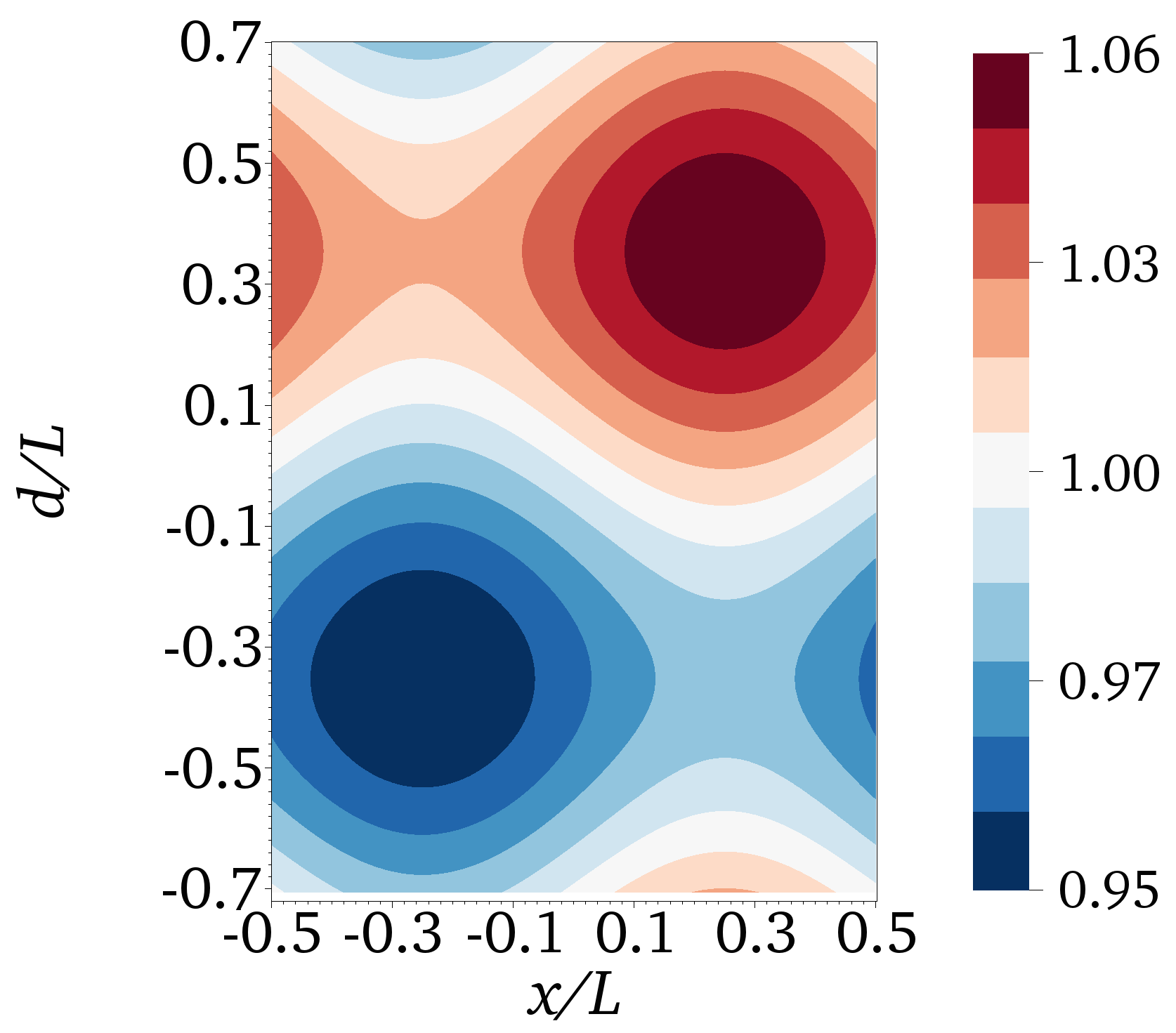}
\includegraphics[width=0.24\textwidth]{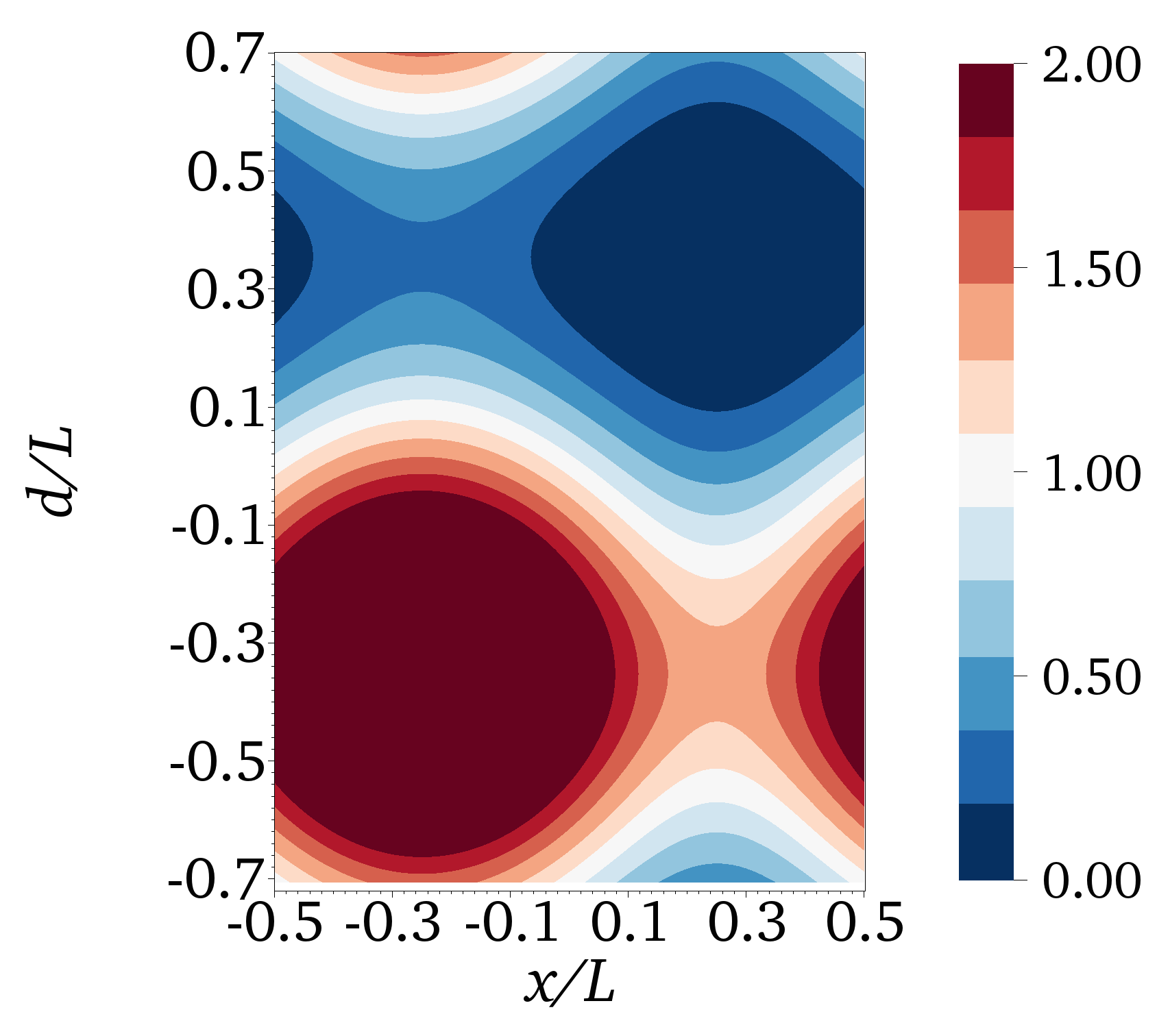}
\caption{
Profile of $\gamma/\bar \gamma$ on the $y=z$ plane ($d=\sqrt{y^2+z^2}$) for $\delta_i=10^{-2}$, 
when $a = a_i$ (left) and when  $a \sim 96 a_i$ (right).}
\label{fig:W}
\efi
\ece
We then evolve the coupled gravitational and hydrodynamical equations,  until the linear size of the domain has increased by roughly
100 times. 
We  measure the
departure of the volume expansion,  represented by $a_{\cal D}$, from the EdS background model, for different 
initial amplitudes of the density contrast $\delta_i$; as clearly shown in Fig.~\ref{fig:V}, this difference is always small. 
We also monitor the density contrast 
at the overdensities and underdensities.
As expected from linear perturbation theory, and shown in Fig.~\ref{fig:d}, 
for small values of the initial $\delta_i$  the density contrast 
grows linearly with $a$, with a well-behaved evolution through $a/a_i = 100$. For  $\delta_i=10^{-2}$,
there is a clear departure from this behavior, with the overdensity becoming nonlinear already at $a/a_i=5$, 
and eventually growing unbounded when $a/a_i \sim 96$.
\bce
\bfi[!hb]
\vspace{0.5mm}
\includegraphics[width=0.4\textwidth]{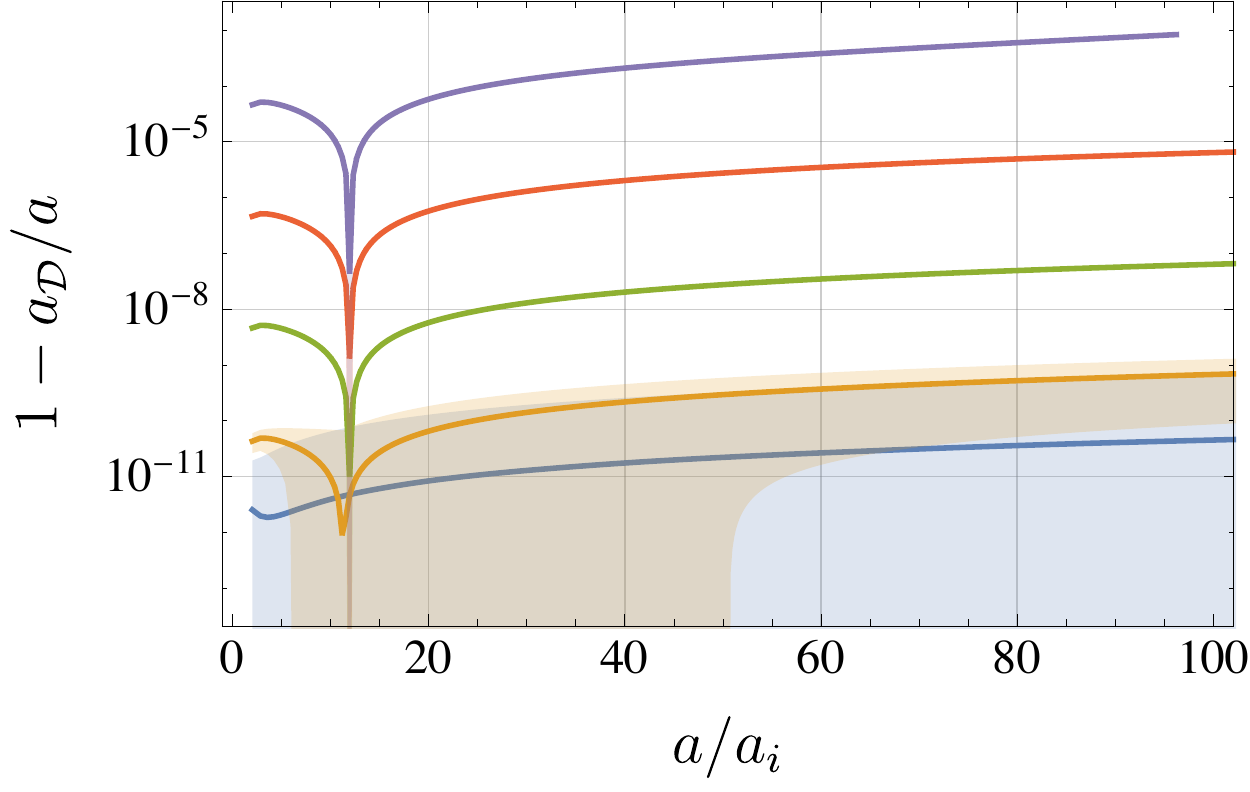}
\caption{
Fractional difference of the scale factor $a_{\cal D}$  of the simulation domain with respect to the 
EdS scale factor $a$,
as a function of the equal-time $a$, for 
for $\delta_i=10^{-2}, 10^{-3}, 10^{-4}, 10^{-5}, 10^{-6}$ (top to bottom).
The numerical error bars, where visible, are included as shaded regions.
\label{fig:V}}
\efi
\ece
\bce
\bfi
\vspace{0.5mm}
\includegraphics[width=0.4\textwidth]{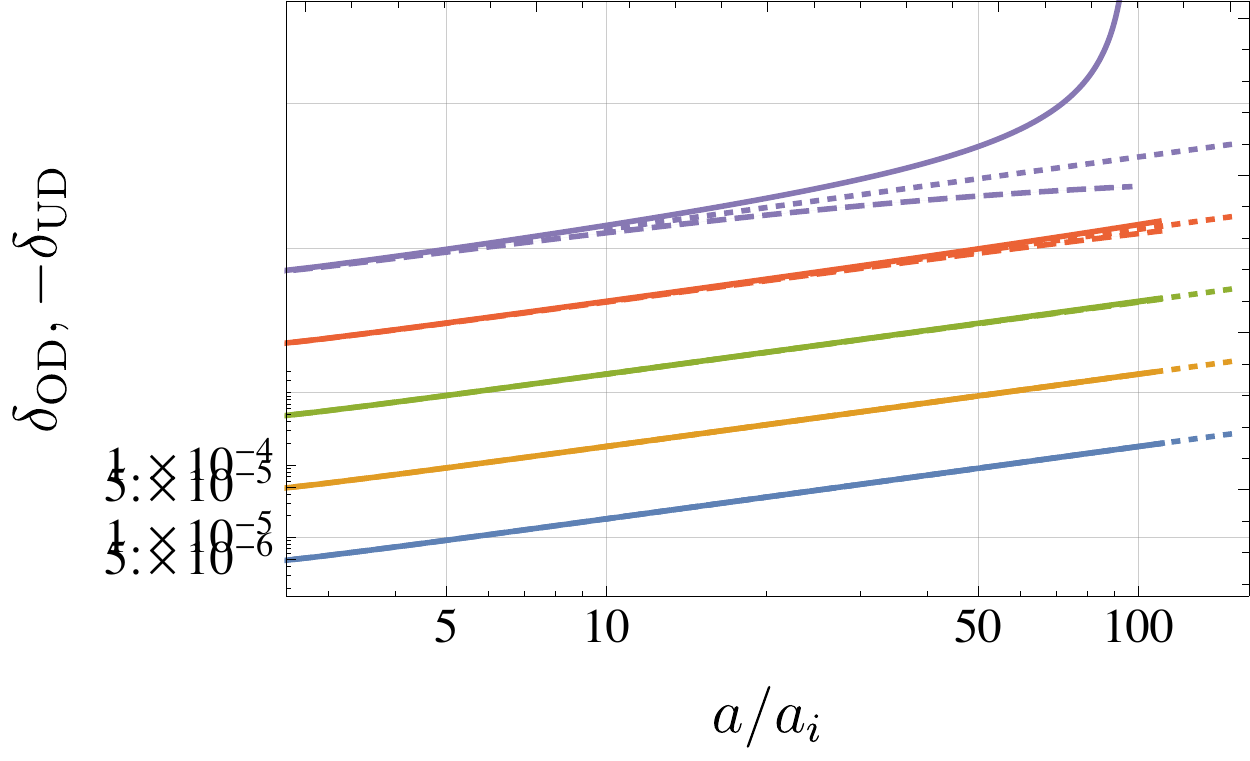}
\caption{
Growth of the density contrast $\delta_{\rm OD}$ at the overdensities (solid lines) and
its negative $-\delta_{\rm UD}$ at the underdensities (dashed lines), for $\delta_i=10^{-2}, 10^{-3}, 10^{-4}, 10^{-5}, 10^{-6}$ (top to bottom).
The linear-perturbation behavior is indicated by dotted lines.
\label{fig:d}}
\efi
\ece
In Fig.~\ref{fig:K} we plot the fractional difference of $K$ (the local expansion rate) from the background value $\bar K=-3 H$ 
at the overdensities
and underdensities.
As expected, the expansion is larger at the underdensities and smaller at the overdensities.
 For $\delta_i=10^{-2}$,
the departure from the expansion rate of the EdS background is substantial: again, the expansion is 
already visibly nonlinear at $a/a_i=5$, and the overdensity reaches the turnaround point 
(signalled by $K_{\rm OD}=0$) at $a/a_i \sim 60$.
At turnaround, the linearly extrapolated density contrast is only $\delta_T=0.6$, much smaller 
than the standard value from spherical top-hat collapse, $\delta_T=1.06$~\cite{Peacock:1999ye}.
For the same initial density contrast, the underdensity
asymptotically approaches the expansion of the Milne model (a 
vacuum FLRW model with negative spatial curvature, represented by the solid gray line in Fig.~\ref{fig:K}), 
as predicted in~\cite{Bruni:1994nf}, with a fractional departure from EdS of over $28\%$ at 
$a/a_i \sim 96$.
These are the first two important results of our calculations: even in this
simple setup, with a perturbation wavelength initially four times larger than the EdS Hubble horizon, the onset of nonlinearity can occur very early, and inhomogeneities can affect the local expansion rate 
in a substantial, nonperturbative way.

In particular, the observed difference with respect to the spherical homogeneous 
top-hat collapse is due to the interplay of several factors, most notably the 
inhomogeneous character of the density, expansion rate, and 3-curvature, and the 
non-vanishing shear $\sigma$, absent in the top-hat case.
Whilst, in the latter case, the perturbation is constrained to remain spatially constant, 
an inhomogeneous density and expansion accelerate the approach to turnaround at the peak, 
just like they do in the spherical Newtonian case~\cite{Bruni:1994cv,Rubin:2013sfa}. 
The shear also gives a small correction, which for $\delta_i=10^{-2}$ is non negligible
even in the initial perturbative regime~\cite{Bouchet:1992uh, Tellarini:2015faa, Bruni:2013qta}.
These effects combine, leading to a negative contribution to the evolution of the local 
expansion rate $-K$, pushing it towards the turnaround ($K=0$), and accelerating the collapse. 
The difference with the top-hat collapse is an important issue which we will investigate in 
detail in future work. 

\bce
\bfi
\vspace{0.5mm}
\includegraphics[width=0.4\textwidth]{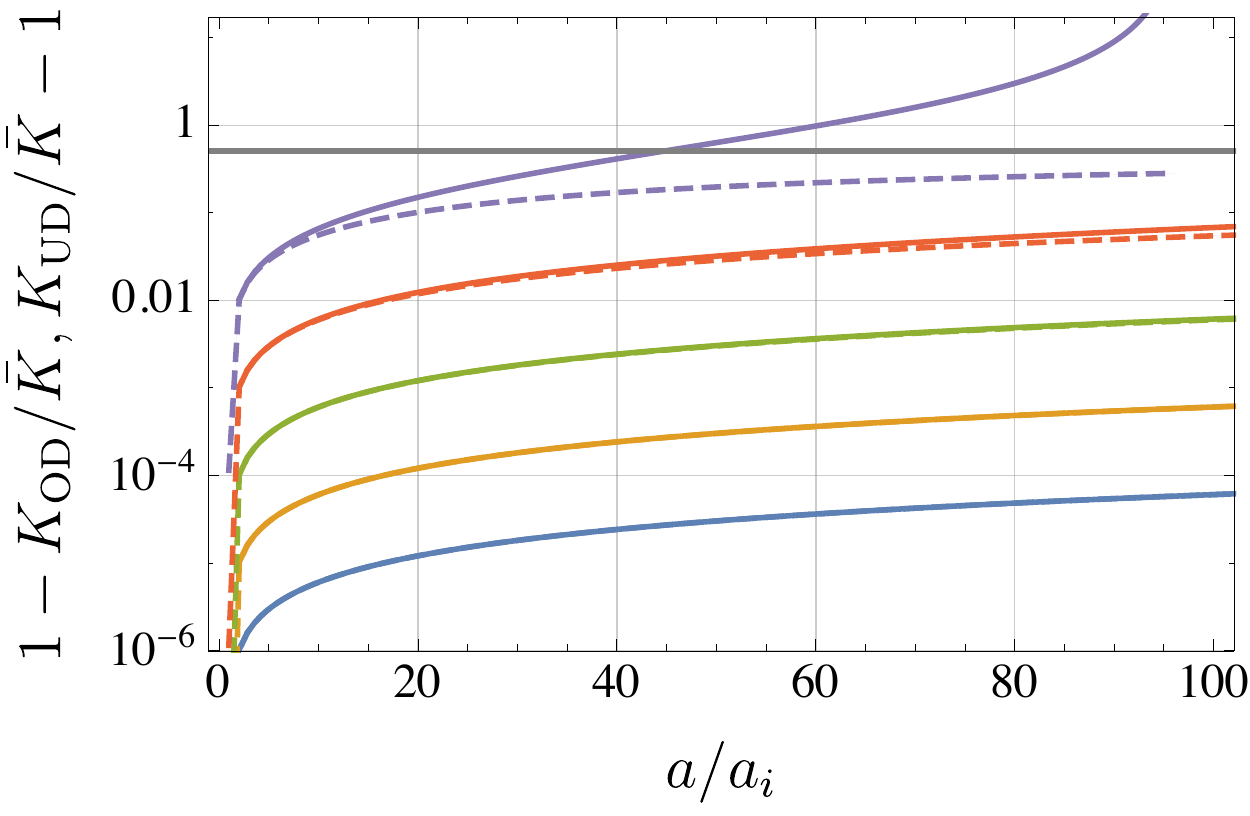}
\caption{
Fractional expansion rate $1-K_{\rm OD}/\bar K$ at the overdensities (solid lines) and
its negative $K_{\rm UD}/\bar K-1$ at the underdensities (dashed lines), for $\delta_i=10^{-2}, 10^{-3}, 
10^{-4}, 10^{-5}, 10^{-6}$ (top to bottom). For $\delta_i=10^{-2}$ the overdensity starts 
collapsing at $a \sim 60 a_i$. The underdensity with $\delta_i=10^{-2}$ expands much 
faster than the background, asymptotically approaching the expansion of the Milne model (horizontal dark-gray line).
\label{fig:K}}
\efi
\ece

We then proceed  to measure the backreaction quantity ${\cal Q}_{\cal D}$: the results are
shown in Fig.~\ref{fig:Q}. 
We extract a few relevant facts: first, given our initial conditions $K_i=\bar{K}_i=-3H_i$, 
it follows from the definition (\ref{eq:br}) that ${\cal Q}_{\cal D}$ vanishes on the initial 
time slice. 
We also notice that, for smaller perturbations, ${\cal Q}_{\cal D}$ remains zero within 
our error bars; for larger perturbations, it is clear from Fig.~\ref{fig:Q} that 
${\cal Q}_{\cal D}$ goes through a short transient phase before following the
scaling ${\cal Q}_{\cal D} \sim a^{-1}$ for a period which is shorter
for higher $\delta_i$.
Given that the only second-order contributions to ${\cal Q}_{\cal D}$ are boundary
terms that vanish on periodic domains like the one we used~\cite{Li:2007ny},
we conjecture that only higher-order terms are contributing to ${\cal Q}_{\cal D}$.
Finally, ${\cal Q}_{\cal D}$ enters
the nonperturbative regime, where it
is negative and its absolute value increases linearly with the scale factor.
The effect is a very small deceleration of the expansion with respect to the EdS model. 
We conclude that the absolute value of ${\cal Q}_{\cal D}$ remains generally quite small, but is not identically zero, 
as would follow from the assumptions of~\cite{Green:2010qy}.

Measuring the sign and scaling of the backreaction ${\cal Q}_{\cal D}$ is a particularly
relevant task, as many speculations on the effect of inhomogeneities on the average
cosmic expansion rate are based on conjectures on these two properties. A back-of-the-envelope
estimate involves the comparison of two competing effects, as quantities like the
matter density at the overdensities quickly depart from the background value,
but at the same time these regions take up a decreasing fractional volume and become 
proportionally less and less relevant to the average.
Our results indicate that, at least for the specific configuration studied here, the
former effect prevails, and the balance is towards an overall slowdown of the expansion rate.

In summary, within the limitations of our setup, in this work we found that, whilst local 
departures from the background density and expansion rate can be tangible,
the average behavior of large volumes remains close to the FLRW background.

\bce
\bfi
\vspace{0.5mm}
\includegraphics[width=0.4\textwidth]{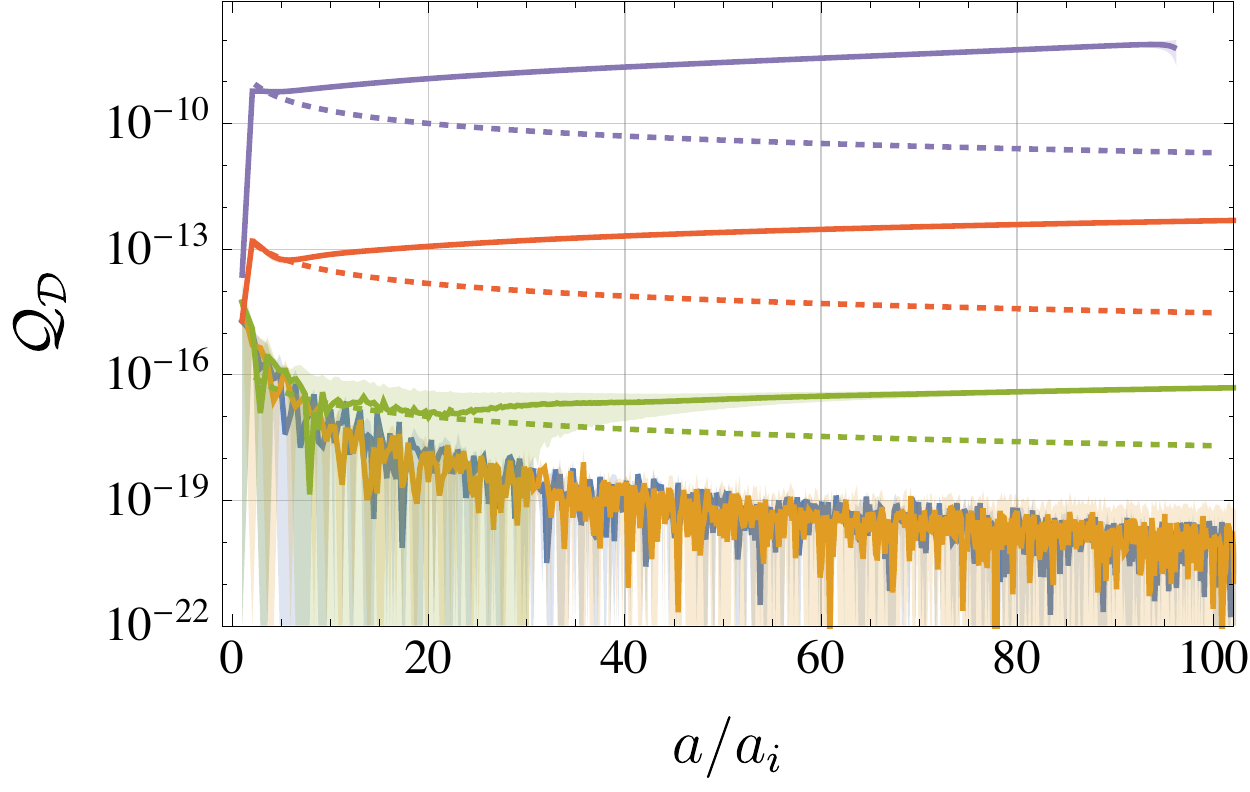}
\caption{
Absolute value of the backreaction ${\cal Q}_{\cal D}$
as a function of the equal-time scale factor in Einstein-de~Sitter space, 
for $\delta_i=10^{-2}, 10^{-3}, 10^{-4}, 10^{-5}, 10^{-6}$ (top to bottom).
The numerical error bars, where visible, are included as shaded regions.
For comparison, we have superimposed dashed lines representing the 
${\cal Q}_{\cal D} \sim a_{\cal D}^{-1}$ scaling. 
\label{fig:Q}}
\efi
\ece

\begin{acknowledgments}
We are grateful to Alessio Notari and an anonymous referee for enlightening comments on this work.
E.B.~is supported by the project
``Digitizing the universe: precision modelling for precision cosmology'', 
funded by the Italian Ministry of Education, University and Research (MIUR). 
M.B.~is supported by the UK STFC Grant No. ST/K00090X/1 and ST/N000668/1. 
The simulations presented in this paper were carried out on the Sciama 
supercomputer at the Institute of Cosmology and Gravitation in Portsmouth.
\end{acknowledgments}

\bibliography{refs}

\end{document}